\documentclass[letterpaper]{article}
\usepackage{iccc}

\usepackage{subcaption}
\usepackage{float}
\usepackage{caption}
\usepackage{times}
\usepackage{helvet}
\usepackage{courier}
\usepackage{graphicx}
\usepackage{amsfonts}
\usepackage{amsmath}
\usepackage{amssymb}
\usepackage{multirow}
\usepackage{hyperref}

\newcommand{\ee}[2]{\mathbb{E}_{#1}\left[#2 \right]}

\title{Neural Drum Machine :\\ An Interactive System for Real-time Synthesis of Drum Sounds}
\author{Cyran Aouameur\\
Sony CSL Paris\\
cyran.aouameur@sony.com\\
\And
Philippe Esling\\
IRCAM \\
esling@ircam.fr
\And
Ga\"etan Hadjeres\\
Sony CSL Paris\\
gaetan.hadjeres@sony.com
}

\begin{document} 
\maketitle
\begin{abstract}
\begin{quote}
In this work, we introduce a system for real-time ge\-neration of drum sounds. This system is composed of two parts: a generative model for drum sounds together with a Max4Live plugin providing intuitive controls on the generative process. The generative model consists of
a Conditional Wasserstein autoencoder (CWAE), which learns to generate Mel-scaled magnitude spectrograms of short percussion samples, coupled with a Multi-Head Convolutional Neural Network (MCNN) which estimates the corresponding audio signal from the magnitude spectrogram.
The design of this model makes it lightweight, so that it allows one to perform real-time generation of novel drum sounds on an average CPU, removing the need for the users to possess dedicated hardware in order to use this system.
We then present our Max4Live interface designed to interact with this generative model. With this setup, the system can be easily integrated into a studio-production environment and enhance the creative process.
Finally, we discuss the advantages of our system and how the interaction of music producers with such tools could change the way drum tracks are composed.

\end{quote}
\end{abstract}

\section{Introduction}

In the early '80s, the widespread use of the sampler revolutionized the way music is produced: besides hiring professional musicians, music producers have since been able to compose with sampled sounds. This has brought much flexibility for both drum and melody production, thanks to the various offline edition possibilities offered by such systems like pitch shifting, time stretching, looping and others.

Nowadays, many producers still rely on samplers for drums production, mainly due to the always-increasing amount of samples libraries available for download. This has helped music production become increasingly accessible, even to newcomers with no or little notion in sound design. However, relying on samples has also some drawbacks. Indeed, producers now have to browse their vast collection of samples in order to find the "right sound". This process is often inefficient and time-consuming. Kick drum datasets are usually unorganized with, for instance, samples gathered in a single folder, regardless of whether they sound "bright" or "dark". As a result, many producers would rely only on a limited selection of their favourite sounds, which could hamper creativity.

Hence, a method allowing a comfortable and rich exploration of sounds becomes an essential requirement in music production, especially for non-expert users. Numerous research efforts have been done in the domain of user experience in order to provide interfaces that enhance the fluidity of human-machine interactions. As an example, synthesizers interfaces now often feature "macro" controls that allow to tune a sound to one's will quickly. 

Another approach to tackle this problem is the use of Music Information Retrieval (MIR) to deal more efficiently with vast libraries of audio samples. MIR is an approach based on feature extraction: by computing a lot of audio features \cite{peeters2004large} over a dataset, one can define a perceptual similarity measure between sounds. Indeed, audio features are related to perceptual characteristics, and a distance between a combination of features is more relevant than a squared error between two waveforms. The combination of MIR with machine learning techniques appears natural in order to organize such audio libraries by allowing, for example, clustering or classification based on audio content. We can cite software such as AudioHelper's Samplism, Sononym and Algonaut's Atlas.

While such software only allows one to organize an existing database, we propose to use artificial intelligence to intuitively generate sounds, thus also tackling the problem of sound exploration. Only very recently, some machine learning models have been developed specifically for the problem of audio generation. These \textit{generative models} perform what we could define as \textit{synthesis by learning}. They rely on generative modelling, which allows performing audio synthesis by learning while tackling the question of intuitive parameter control \cite{esling2018generative,engel2017neural}.

\begin{figure*}[!ht]
\centering
\includegraphics[width =0.9\textwidth]{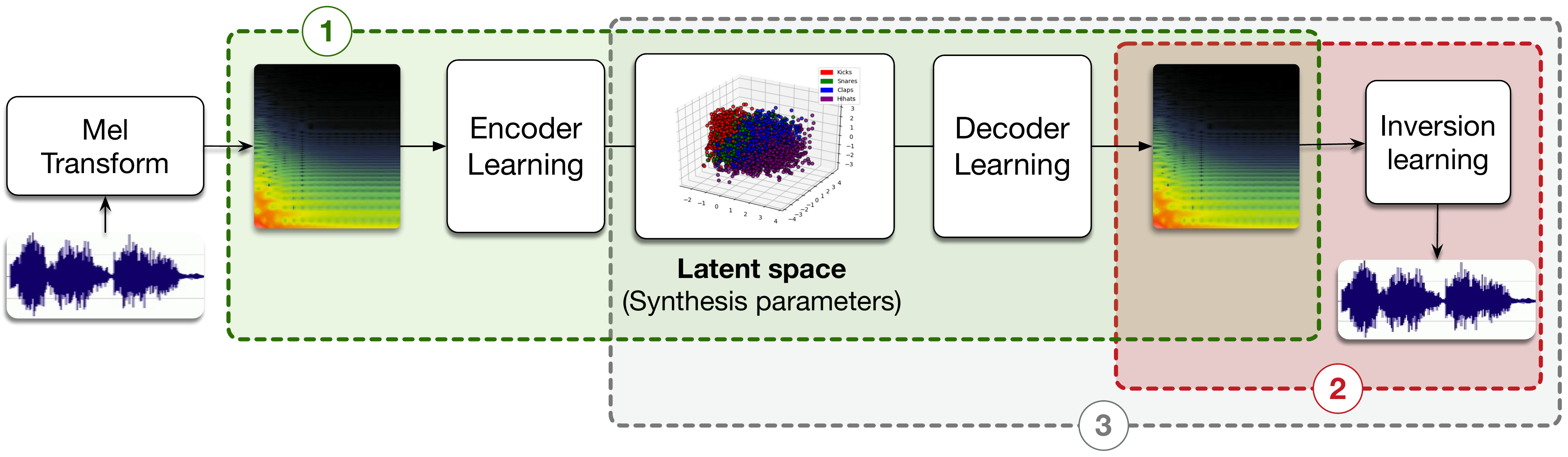}
\caption{This diagram presents our end-to-end system for drum sounds synthesis. The generative model (1) learns how to reconstruct spectrograms from a parameters' space. Then, the second part of the system (2) is dedicated to spectrogram inversion, to generate some signal from a Mel spectrogram. Finally, the software interface (3) allows a user to interact with the model and to generate sound from the parameters' space.}
\label{fig:workflow}
\end{figure*} 

\textit{Generative models} are a flourishing class of machine learning approaches whose purpose is to generate novel data based on the observation of existing examples \cite{bishop2014pattern}. The learning process consists of modelling the underlying (and unknown) probability distribution of the data based on samples. 

Once the model is trained, it is then possible for a user to generate new samples at will. However, for the user to be active during the synthesis process and not only passively browsing the outputs of the system, we find crucial the requirement that the system should provide intuitive controls. To this end, we need a model that extracts a compact high-level representation of the data. Then, by providing these simple high-level controls to a user, the synthesis process can be guided by perceptual characteristics. A user would just have to explore a continuous and well-organized parameter space to synthesize an infinite variety of sounds.

\subsection{Our proposal}
\label{sec:proposal}

In this work, we describe a system that allows to create a controllable audio synthesis space so that we can use it to synthesize novel sounds in an intuitive manner. This system can be split  into three components (Fig.~\ref{fig:workflow}):

\begin{itemize}
    \item A Conditional Wasserstein Auto-Encoder (CWAE) which generates Mel-scaled spectrograms.
    \item An extension of the Multi-Head Convolutional Neural Network (MCNN) which reconstructs signal from Mel-scaled spectrograms.
    \item A Max4Live plugin allowing users to interact with the model in a music production environment.
\end{itemize}

In the remainder of this document, we first provide a state of the art on Wasserstein auto-encoders and MCNN. Then we describe our model and the data we used to train it. We discuss reconstruction and generation results. Finally, we showcase the associated plugin and explain how it could change the way drum tracks are produced.

\section{Related work}

\subsection{Generative models on audio waveforms}

A few systems based on generative models have been recently proposed to address the learning of latent spaces for audio data.
The Wavenet auto-encoder \cite{engel2017neural} combines Wavenet \cite{oord2016wavenet} with auto-encoders and uses dilated convolutions to learn waveforms of musical instruments. By conditioning the generation on the pitch, such a system is capable of synthesizing musical notes with various timbres. The WaveGAN \cite{donahue2018adversarial} uses Generative Adversarial Networks (GANs) to generate drum sounds or bird vocalizations by directly learning on waveform. However, the GAN approach provides little control over the generation because it is still difficult to structure their latent space. %ref ?

\subsection{Generative models on spectral representations}
Other works have focused on generating sound as spectrograms, a complex time-frequency representation of sound. This visual representation of sound intensity through time allows us to treat sounds like images, but has to reverted back to the signal domain to produce sound.
In \cite{esling2018generative} uses VAEs to learn a generative space where instrumental sounds are organized with respect to their timbre. However, because the model is trained on spectra frames, it lacks temporal modeling. This hampers the capacity of the model to easily allow users to generate evolving structured temporal sequences such as drum sounds. 
This approach introduced in \cite{donahue2018adversarial} takes into account these temporal dependencies by proposing SpecGAN, a generative models which uses GANs to generate spectrograms as if they were images.

\subsection{Spectrogram inversion}
Working with neural networks often forces us to discard the phase information of a spectrogram. Therefore, one cannot use the inverse Fourier transform to retrieve the signal it originates from. With classic STFT, a common workaround is to use the Griffin-Lim Algorithm (GLA) \cite{griffin1984signal} which allows to estimate the missing phase information.
Also, The Multi-head Convolutional Neural Network (MCNN) is a model that inverts STFTs \cite{arik2019fast} using neural networks.

However, STFT is not the best transform for our purpose. Indeed, Mel-scaled spectrograms are known to be more suitable for training convolutional neural networks \cite{huzaifah2017comparison}. Mel-scaled spectrograms are computed with filters based on the Mel scale, a perceptual frequency scale that tries to mimic the human perception of pitches.

Despite being more adapted for training, using Mel-scaled spectrograms introduces a problem: they are not invertible and GLA cannot be used. 
Therefore, some deep learning based models have been developed in order to estimate signal from non-invertible spectrograms.  In \cite{prenger2018waveglow}, the authors present WaveGlow, a flow-based network capable of generating high quality speech from Mel spectrograms. Also, in \cite{huang2018timbretron}, the authors use a conditioned Wavenet to estimate signal from Constant-Q Transforms, another non-invertible transform.

\section{Proposed model}
Our model is composed of two components: a generative model on spectrograms, whose role is to learn a latent space from our dataset and to generate meaningful spectrograms from this space, and a spectrogram inversion model, whose role is reconstruct waveforms from our generated spectrograms.

\subsection{Preliminaries on variational autoencoders}
To formalize our problem, we rely on a set of data $\{\mathbf{x}_{n}\}_{n\in[1,N]}$ lying in a high-dimensional space $\mathbf{x}_i\in\mathbb{R}^{d_{x}}$. We assume that these examples follow an underlying probability distribution $p\left(\mathbf{x}\right)$ that is unknown.
Our goal is to train a generative model able to sample from this distribution.

We consider a parametrized \textit{latent variable model}
\begin{equation*}
p_\theta(\mathbf{x}, \mathbf{z}) = p_\theta(\mathbf{x} | \mathbf{z})\pi(\mathbf{z}).
\end{equation*}
by introducing latent variables $\mathbf{z}\in\mathbb{R}^{d_{z}}$ lying in a space of smaller dimensionality than $x$ ($d_{z} \ll d_{x}$) and distributed according to the prior $\pi(z)$. We are interested in finding the parameter $\theta$ that maximizes the likelihood $\sum_i p_\theta(x_i)$ of the dataset. However, for usual choices of the conditional probability distributions $p_\theta(x|z)$ (typically a deep neural network), this quantity is intractable.

The variational autoencoder (VAE) \cite{Kingma2013Auto-EncodingBayes} is a model that introduces a variational approximation $q_\phi(\mathbf{z|x})$ to the intractable posterior $p_\theta(\mathbf{x|z})$ (the approximate posterior $q_\phi(\mathbf{z|x})$ is often chosen as a parametrized family of diagonal Gaussian distributions). The network $q_\phi(z|x)$ is called the \emph{encoder} whose aim is to produce latent codes given $x$ while the network $p_\theta(x|z)$ is called a \emph{decoder}, which tries to reconstruct $x$ given a latent code $z$.

The introduction of the variational approximation of the posterior allows us to obtain the following lower bound $\mathcal{L}(\boldsymbol{\theta, \phi})$ (called ELBO for Evidence Lower BOund) over the intractable likelihood:

\begin{multline}
\label{eq:param-obj}
\mathcal{L}(\boldsymbol{\theta, \phi}) = \mathbb{E}_{\mathbf{x} \sim p(\mathbf{x})} \big[ \underbrace{\mathbb{E}_{\mathbf{z} \sim p(\mathbf{z|x})} \big[ \log{ p_\theta (\mathbf{x|z}) } \big]}_{\text{reconstruction}} \\
 - \underbrace{D_\mathrm{KL} \big[ q_\phi(\mathbf{z|x}) \parallel \pi(\mathbf{z}) \big]}_{\text{regularization}} \big],
\end{multline}
where $D_\mathrm{KL}$ denotes the Kullback-Leibler divergence \cite{cover2012elements}.
\begin{itemize}
\item The first term $\mathbb{E}_{\mathbf{z} \sim p(\mathbf{z|x})} \big[ \log{ p_\theta (\mathbf{x|z}) } \big]$ is the likelihood of the data $\mathbf{x}$ generated from the set of latent variable $\mathbf{z} \sim q_\phi(z|x)$ coming from the approximate posterior. Maximizing this quantity can be seen as  minimizing a \textit{reconstruction error}.
\item The second term $D_{KL} \big[ q_\phi(\mathbf{z|x}) \parallel \pi(\mathbf{z}) \big]$ is the distance between $q_\phi(\mathbf{z|x})$ and $\pi(\mathbf{z})$ and can be interpreted as a regularization term.
\end{itemize}  

In \cite{sohn2015learning}, the authors add a conditioning mechanism to the original VAE which consists in conditioning all three networks $p_\theta(x|z)$, $q_\phi(z|x)$ and $\pi(z)$ on some metadata $m$ (in most cases, the prior $\pi(z)$ does not depend on $m$).

However, a known problem of VAEs is that they tend to generate blurry samples and reconstructions \cite{chen2016variational}. This becomes a major hindrance in the context of spectrogram reconstructions. 
Hopefully, this problem can be overcome by the use of Wasserstein Auto-Encoders (WAEs) instead of VAEs. The main difference consists in replacing the $D_\mathrm{KL}$ term in (\ref{eq:param-obj}) by another divergence between the prior $\pi$ and the \textit{aggregated posterior} $q_Z(\mathbf{z}):= \ee{x\sim p_X}{q(\mathbf{z|x})}$.
In particular, the MMD-WAE considers a Maximum Mean Discrepancy (MMD) \cite{berlinet2011reproducing} distance defined as follows:
\begin{equation}
  \label{eq:mmdk}
  \mathrm{MMD}_k^2(p, q) = \bigl\| \int_{Z} k(z, \cdot) p(z)dz - \int_{Z} k(z, \cdot) q(z)dz\bigr\|^2_{\mathcal{H}_k},
\end{equation}
where $k: \mathcal{Z} \times \mathcal{Z} \to \mathbf{R}$ is an positive-definite reproducing kernel and $\mathcal{H}_k$ the associated Reproducing Kernel Hilbert Space (RKHS) \cite{berlinet2011reproducing}.
MMD is known to perform well when matching high-dimensional standard normal distributions \cite{tolstikhin2017wasserstein,gretton2012kernel}.
Since the MMD distance is not available in closed form, we use the following unbiased U-statistic estimator \cite{gretton2012kernel} for a batch size $n$ and a kernel $k$:
\begin{multline}
\label{eq:discrete-MMD}
    \mathrm{MMD}_{k, n}^2(\pi, q_z) := \frac{1}{n(n-1)}\sum_{l \neq j} k(z_l, z_j)  \\
    +\frac{1}{n(n-1)}\sum_{l \neq j} k(\tilde{z}_l, \tilde{z}_j) 
    - \frac{2}{n^2}\sum_{l,j}k(z_l,\tilde{z}_j),
\end{multline}
with $\tilde{z} := \{\tilde{z}_1, \dots, \tilde{z}_n\}$ where $\tilde{z_i} \sim \pi$ and $z := \{z_1, \dots, z_n\}$ where $z_i \sim q_z$.

\subsection{The Conditional WAE}
%Global structure
We now introduce a Conditional WAE (CWAE) architecture so that we can generate spectrograms depending on additional metadata such as the category of the original sound (e.g. kick drum, snare, clap, etc.).
%To learn a latent space and to generate spectrograms, we use a WAE with convolutional layers and add a conditioning mechanism.

Our encoder is defined as a Convolutional Neural Network (CNN) with $l$ layers of processing. Each layer is a 2-dimensional convolution followed by conditional batch normalization \cite{perez2017learning,perez2018film} and a ReLU activation. This CNN block is followed by Fully-Connected (FC) layers, in order to map the convolution layers activation to a vector of size $d_{z}$ which is that of the latent space. The decoder network is defined as a mirror to the encoder, so that they have a similar capacity. Therefore, we move the FC block before the convolutional one and change the convolution to a convolution-transpose operation. Also, we slightly adjust the convolution parameters so that the output size matches that of the input. 

Our convolutional blocks are made of 3 layers each, with a kernel size of (11,5), a stride of (3,2) and a padding of (5,2). Our FC blocks are made of 3 layers with sizes 1024, 512 and $d_{z} = 64$. Therefore, our latent space is of size $d_{z} = 64$. 

In the case of WAEs, the MMD is computed between the prior $\pi$ and the aggregated posterior $q_Z(\mathbf{z}):= \ee{\mathbf{x}\sim p_X}{q(\mathbf{z|x})}$. As a result, the latent spaces obtained with WAEs are often really Gaussian which makes them easy to sample. Here, the conditioning mechanism implies that we use separated gaussian priors $\pi_c = \mathcal{N}(0,1)$ for each class $c$, in order to be able to sample all classes as Gaussian. Indeed, computing a MMD loss over all classes would force the global aggregated posterior to match the gaussian prior, and thus restrict the freedom for latent positions. Therefore, we have to compute the per-class MMD to backpropagate on.

Let's formalize this problem by decomposing our dataset $\mathbb{D}$ into $\mathbb{C}$ subsets $\mathbb{D}_c$ with $1\le c \le \mathbb{C} $, containing all elements from a single class. We define $q^c_z(\mathbf{z}) := \ee{x\in \mathbb{D}_c}{q(\mathbf{z|x},m=c)}$. Thus, our regularizer is computed as follows :
\begin{equation}
  \label{eq:mmdkc}
  \mathcal{D}_Z(\pi_c, q_z) = \frac{1}{\mathbb{C}}\sum_{c=1}^{\mathbb{C}} \mathrm{MMD}^2_{k,n}(\pi, q_z^c).
\end{equation}

Finally, our loss function is computed as:

\begin{equation}
    \mathcal{L}(\boldsymbol{\theta, \phi})  = \sum^n_{i=1} \mathrm{MSE}(x_i, \hat{x_i}) + \beta \mathcal{D}_Z(\pi, q_z),
\end{equation}
where $\beta=10$ and $k$ is the \textit{multi-quadratics kernel} as for CelebA in \cite{tolstikhin2017wasserstein}.

\subsection{MCNN inversion}

\begin{figure}[t]
\centering
\includegraphics[width =0.5\textwidth]{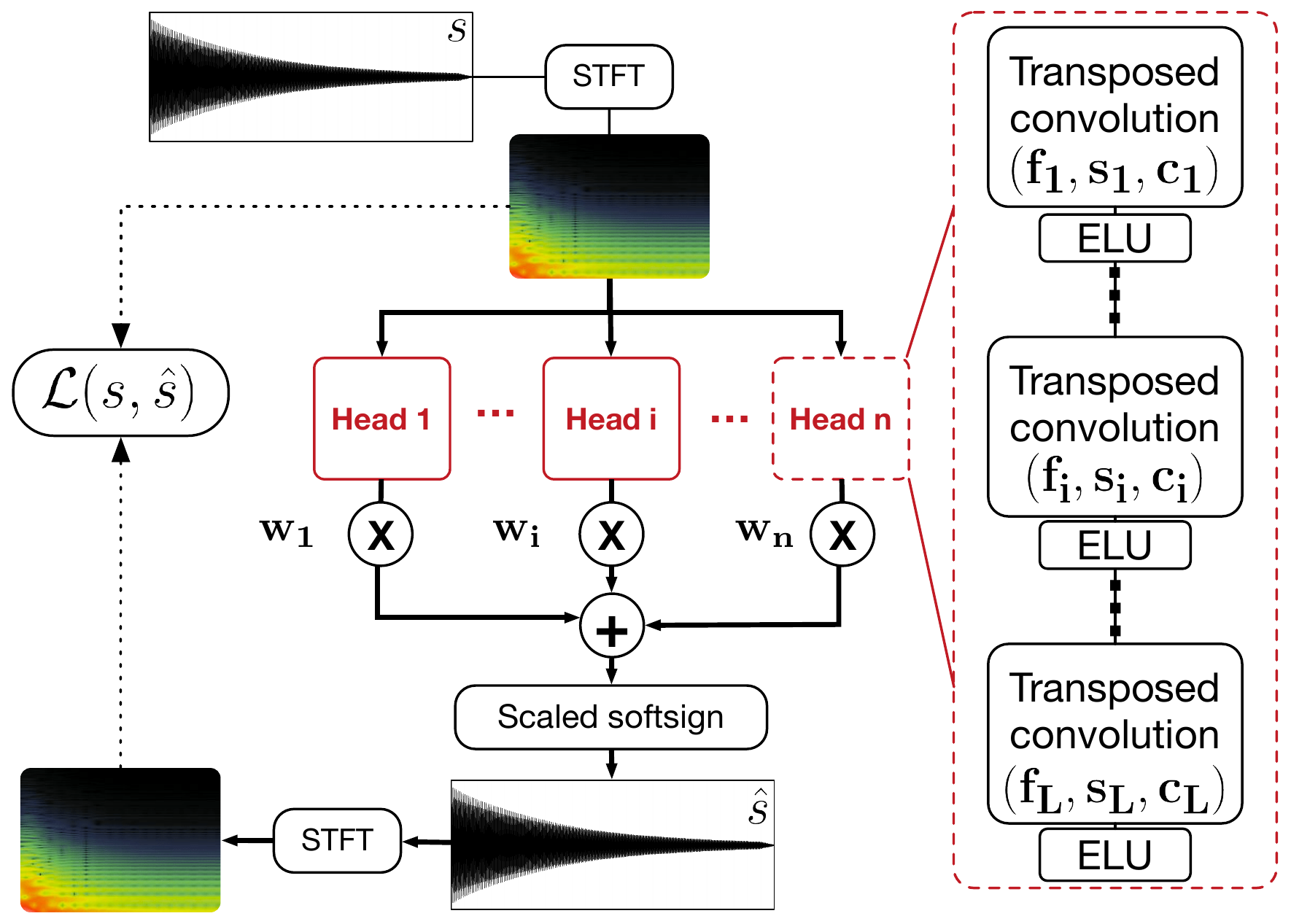}
\caption{The MCNN for spectrogram inversion. Its multiple heads estimate waveforms that are summed to produce the final waveform. Finally, the loss is computed between the resulting spectrogram and the original one}
\label{fig:MCNN}
\end{figure}

To invert our Mel-spectrograms back to the signal domain, we use a modified version of the original MCNN. In this section, we first review the original MCNN before detailing how we adapted it to handle Mel-spectrograms of drum samples.

MCNN is composed of multiple heads that process STFTs (Fig.~\ref{fig:MCNN}). These heads are composed of $L$ processing layers combining 1D transposed convolutions and Exponential Linear Units (ELUs). The convolution layers are defined by a set of parameters $(f,s,c)$, respectively the filter width, the stride and the number of output channels. We multiply the output of every head with a trainable scalar $w_i$ to weight these outputs, and we compute the final waveform as their sum. Lastly, we scale the waveform with a non-linearity (scaled softsign). The model is trained to estimate a waveform which spectrogram matches the original one. For more implementation details, we refer the interested readers to the original article.

We have chosen to use this model because of three main points. First, it performs a fast (300x real-time) and precise estimation of a signal given a spectrogram. Then, it can deal with non-invertible transforms that derive from STFT such as Mel-STFT. Finally, its feed-forward architecture allows to takes advantage of GPUs, unlike iterative or auto-regressive models.

In our implementation, we kept most of the parameters suggested in \cite{arik2019fast}.We use a MCNN with 8 heads of $L=8$ layers each where, for each layer $l_i$, $1 \le i \le L$, we have $(w_i,s_i) = (13, 2)$. However, because we have padded our signals with zeros to standardize their length, two problems appear. 
First, we observed that the part of the spectrogram corresponding to the padding (made of zeros) was not well reconstructed if convolution feature biases. Without biases, zeros stay zeros throughout the kernel multiplications. Therefore, we removed all biases. 
Then, we observed a leakage phenomenon: because the convolution filters are quite large (length 13), the reconstructed waveform had more non-zero values than the original one. 
Therefore, the loss is lower-bounded by this effect. To tackle this problem, we decided to apply a mask to the final output of our model, aiming at correcting this leakage. Thus, for the number of output channels for layer $i$, we have :
\[
  c_i =
    \begin{cases}
    2^{L-i} & \text{if } 2 \le i \le L \\
    2 & \text{if } i = 1.
    \end{cases}
\]
The output of head h is a couple of 2 vectors $(s_h, m_h)$. We estimate the mask $\hat{M}$ as follows:
\begin{equation}
    \hat{M} = \sigma \left (\sum_{h=1}^{8} m_h \right).
\end{equation}

The finally output waveform $\hat{s}$ is computed as :
\begin{align}
    \hat{s}^* & = \sum_{h=1}^{8} w_h * s_h, \\
    \hat{s}  & = \hat{s}^* \times \hat{M}.
\end{align}

To train the mask, we use supervised training and introduce a loss term between the original mask $M$ and the estimated one $\hat{M}$, that we name \textit{mask loss}:
\begin{equation}
\label{eq:ML_eq}
    \textrm{L}_{mask} (M,\hat{M}) = 
    \textrm{BCE}(M,\hat{M}).
\end{equation}
At generation time the mask is binarized. This solution has worked very well to cut the tail artifacts introduced by the convolutions.

A second change is that we now train MCNN on Mel-scaled spectrograms rather than STFT. However, original losses were computed on STFT. To turn a STFT into a Mel-scaled spectrogram, we compute a filterbank matrix $F$ to combine the 2048 FFT bins into 512 Mel-frequency bins. Finally, we multiply this matrix with the STFT to retrieve a Mel-scaled spectrogram: 
\begin{equation}
\label{eq:stft2mel}
    \textrm{Mel} = \textrm{STFT} \times F.
\end{equation}
Therefore, we can simply convert all STFTs to Mel-scaled spectrograms before the loss computation. This does not affect the training procedure: back-propagation remains possible since this conversion operation is differentiable.

In addition, we have modified the loss function. When training the original model on our data, we noticed some artifacts that we identified as 'checkerboard artifacts'. These are known to appear when using transposed convolutions \cite{odena2016deconvolution}. We have tried known workarounds such as NN-Resize Convolutions \cite{aitken2017checkerboard} but it did not yield better results. We empirically realized that, in our particular case, removing the phase-related loss terms helped reducing these artifacts.
Therefore, we removed from \cite{arik2019fast} the instantaneous frequency loss and the weighted phase loss terms while keeping the Spectral Convergence (SC) term:

\begin{equation}
\label{eq:SC_eq}
    \textrm{SC} (s,\hat{s}) = 
    \frac{\| |\textrm{MEL}(s)| - |\textrm{MEL}(\hat{s})| \|_{F}}{\||\textrm{MEL}(s)|\|_{F}},
\end{equation}
where $\|\cdot\|_{F}$ is the Frobenius norm over time and frequency, and the Log-scale MEL-magnitude loss ($\textrm{SC}_{log}$):
\begin{equation}
\label{eq:log_MEL_eq}
    \textrm{SC}_{log} (s,\hat{s}) = 
    \frac{\|\log (|\textrm{MEL}(s)| + \epsilon) - 
    \log(|\textrm{MEL}(\hat{s})| + \epsilon)\|_1}{\log(|\textrm{MEL}(s)| + \epsilon)\|_1} ,
\end{equation}
where $\|\cdot\|_1$ is the $L^1$ norm and $\epsilon$ is a small number.

\vspace{0.01\textwidth}
Finally, our global loss term is:
\begin{equation}
\label{eq:totalloss}
    L= \alpha \textrm{SC}(s,\hat{s})+ \beta \textrm{SC}_{log}(s,\hat{s})+ \gamma\textrm{L}_{mask}(M,\hat{M}),
\end{equation}
where $\alpha,\beta$ and $\gamma$ are constants used for weighting loss terms. In our experiments, we set $(\alpha, \beta, \gamma) = (3,10,1) $, which works well in practice.

\section{Experiments}

\subsection{Dataset}
\label{sec:dataset}
We built a dataset of drums samples coming from various sample packs that we have bought (Vengeance sample packs and others). Overall, we collected more than 40,000 samples across 11 drum categories. All sounds are WAV audio files PCM-coded in 16 bits and sampled at 22050 Hz. Sounds that were longer than 1 second were removed in order to obtain a homogeneous set of audio samples. 

After this preprocessing, the final dataset contains 11 balanced categories (kicks, claps, snares, open and closed hi-hats, tambourines, congas, bongos, shakers, snaps and toms) with 3000 sounds each for a total of 33000 sounds. All sounds in the dataset have a length between 0.1 and 1 second (mean of 0.46 second). In order to validate our models, we perform a class-balanced split between 80\% training and 20\% validation sets. All the results we present are computed on this validation set to ensure generalization.

As said in previous sections, we compute the Mel-scaled spectrograms of these sounds. To do so, we first pad all waveforms with zeros to ensure a constant size among the whole dataset. Thus, all audio files are 22015 samples long. We also normalize them so that the maximum absolute value of samples is 1. Then, we compute STFTs for all sounds with a Hann window with a length of 1024, a hop size of 256 and an FFT size of 2048. To turn the STFTs into Mel-scaled spectrograms, we multiply the STFTs with the filter-bank matrix we mentioned earlier (Eq.~\ref{eq:stft2mel}).

\begin{figure*}[]
\centering
\begin{subfigure}{0.4\textwidth}
\centering
\includegraphics[width=\columnwidth]{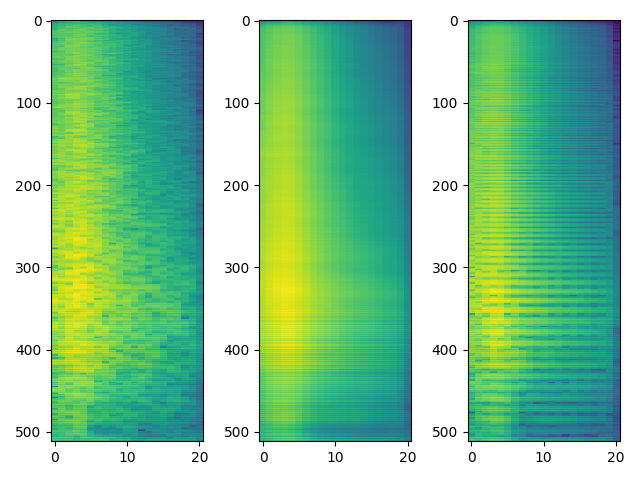}%
\caption{Clap}%
\label{fig:clapsrec}%
\end{subfigure}%
\hspace{0.1\textwidth}
\begin{subfigure}{0.4\textwidth}
\centering
\includegraphics[width=\columnwidth]{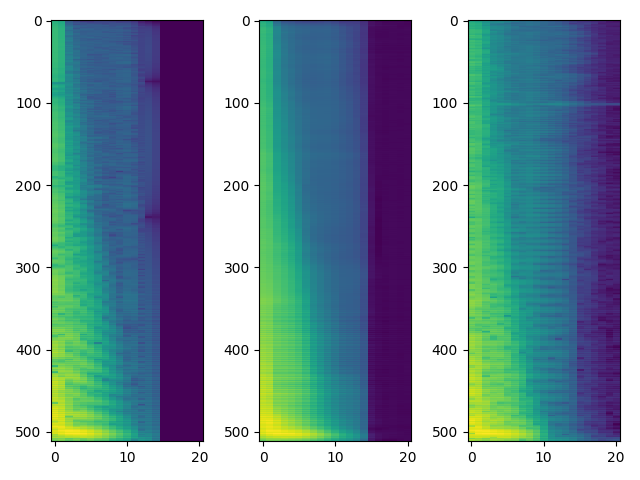}%
\caption{Kick drum}%
\label{fig:kicksrec}%
\end{subfigure}
\caption{Spectrogram reconstructions of sounds from the evaluation set. From left to right, we have: the original spectrogram, the CWAE reconstruction and the one obtained from the reconstructed waveform (the amplitudes are presented in log-scale for the sake of visibility}
\label{fig:specrec}
\end{figure*}

\subsection{Experimental setup}

Before assembling the two parts of our model to create an end-to-end system, we pre-train each network separately. 

We train our CWAE with an ADAM optimizer \cite{kingma2014adam}. The initial learning rate is set to $\eta = 1e^{-3}$ and is annealed whenever the validation loss has not decreased for a fixed number of epochs. The annealing factor is set to 0.5 and we wait for 10 epochs. The WAE is trained for 110k iterations. To obtain a good estimation of the MMD between each $q_Z^c$ and their Gaussian prior, we have to compute enough $z$. Indeed, it is said in \cite{reddi2015high} that $n$ in equation~\ref{eq:discrete-MMD} should be the same order of magnitude as $d_z$ = 64. Therefore, at each iteration, we have to ensure that this criterion is satisfied for each class. We then implemented a balanced sampler, for our data loader to yield balanced batches containing 64 samples for each class. It ensures more stability than a standard random batch sampler. In the end, our final batch size equals $64 \times 11 = 704$.

When training the CWAE, we perform some data processing steps that allow greater stability and performance. First, we compute the log of our spectrograms to reduce the contrast between high and low amplitudes. Then, we compute the per-element means and variances to scale the log-Mel spectrograms so that each element is distributed as a zero-mean unit-variance Gaussian. Indeed, we have noticed that it improves the WAE reconstruction quality. 

When training the MCNN, we use the Mel spectrograms without scaling. The initial learning rate is set to $\eta = 1e^{-4}$ and is annealed by a scheduler at a rate of 0.2 with a patience of 50 epochs. The MCNN is trained for around 50k iterations, with a batch size of 128.

\subsection{Reconstruction}

\begin{figure}[]
\centering
\begin{subfigure}{0.5\columnwidth}
\centering
\includegraphics[width=\columnwidth]{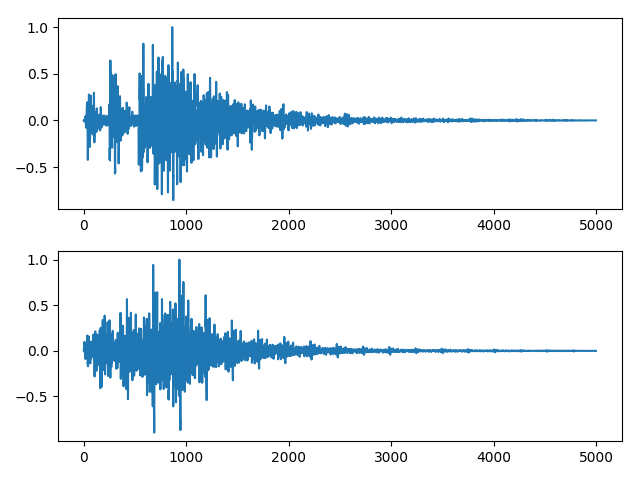}%
\caption{Clap}%
\label{fig:clapwrec}%
\end{subfigure}%
\hfill
\begin{subfigure}{0.5\columnwidth}
\centering
\includegraphics[width=\columnwidth]{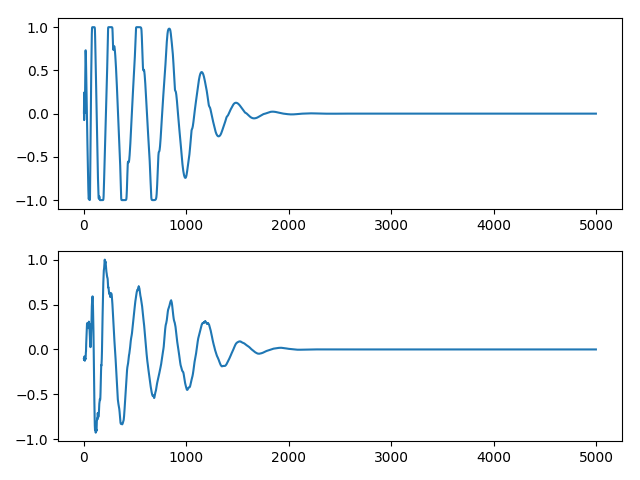}%
\caption{Kick drum}%
\label{fig:kickwrec}%
\end{subfigure}
\caption{Waveform reconstruction of sounds from the evaluation set. The top row shows the original waveform and the bottom shows the reconstruction after passing the spectrogram throughout the whole system}
\label{fig:waverec}
\end{figure}

We first evaluate the reconstruction abilities of each part of our system, and the system as a whole. On figure~\ref{fig:specrec}, we compare the original spectrogram with both our CWAE's reconstruction and the spectrogram computed on the final output. In both cases, the reconstruction performed by the CWAE is good yet a bit blurry. After passing through the MCNN, we can see some stripes, corresponding to some checkerboard artifact, which periodically affects the waveform. Thus, this appears as a harmonic artifact on the spectrogram. While appearing important on these spectrograms because of the log, the sound is often clean, as shown on the kick reconstruction on figure~\ref{fig:waverec}.

More examples are available on the companion website\footnote{https://sonycslparis.github.io/NeuralDrumMachine/}, along with audio.
\begin{figure*}[!ht]
\centering
\includegraphics[width =\textwidth]{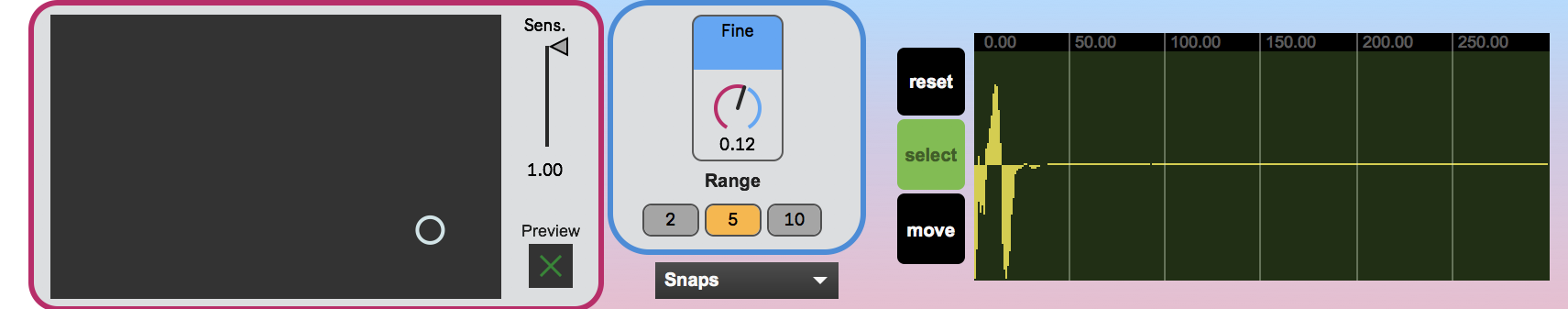}
\caption{The Neural Drum Machine interface. First, the XY pad on the left controls values for the two most influential dimensions. The "Fine" knob controls the value for the third most influential dimension and can be seen as fine tuning. The range selector controls the range of values available for these three dimensions.
Then, a selector allows the user to control which type of sound is generated. Finally, the waveform visualizer on the right allows to trim a sample to play only a particular region.}
\label{fig:ndm}
\end{figure*}

\subsection{Sampling the latent space}

\begin{figure}[h!]
\centering
\begin{subfigure}{0.4\columnwidth}
\centering
\includegraphics[width=\columnwidth]{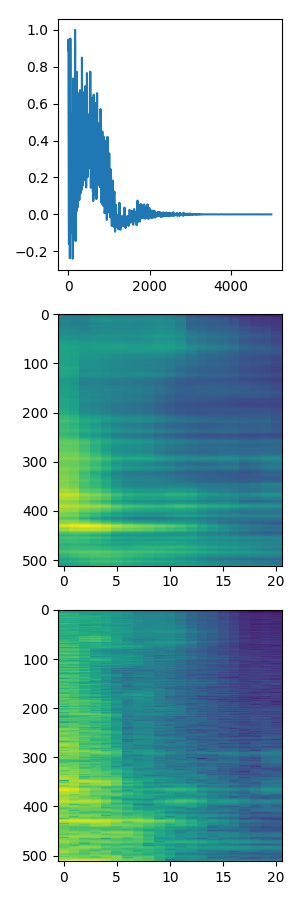}%
\caption{Bongo}%
\label{fig:genbongo}%
\end{subfigure}%
\hspace{0.1\columnwidth}
\begin{subfigure}{0.4\columnwidth}
\centering
\includegraphics[width=\columnwidth]{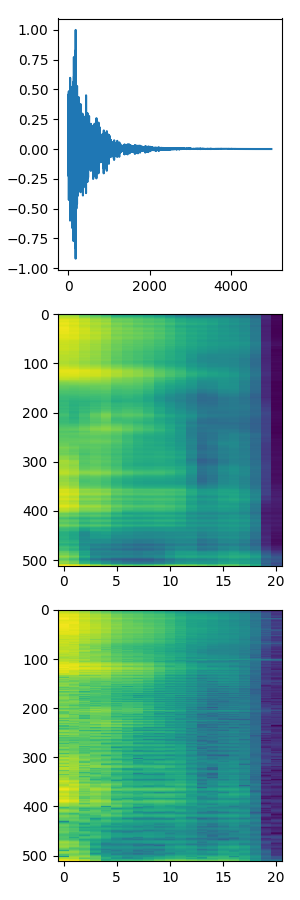}%
\caption{Hi-hat}%
\label{fig:genhh}%
\end{subfigure}
\caption{Sounds generated by sampling the latent space. From top to bottom, we have the final waveform, the spectrogram generated by the CWAE and the one corresponding to the waveform (the amplitudes are presented in log-scale for the sake of visibility).}
\label{fig:gen}
\end{figure}

On figure~\ref{fig:gen}, we show generated sounds. We generate them by first sampling a multivariate Gaussian in the latent space. Then, we decode this latent code, conditioned on a given class label and obtain a spectrogram. Finally, this spectrogram is passed to the MCNN which estimates the corresponding waveform. Here, both these sounds are pretty realistic and artifact free. However, sampling the latent space in this fashion does not always yield good sounding results. This is because our latent distributions do not really match Gaussian distributions. Also, conditioning on a category does not ensure to generate sounds from this category only. Indeed, some regions of the space will sound close to a hi-hat, even if the class label for claps, is provided to the CWAE. While this can be seen as a drawback, we think that this does not lower the interest because it allows synthesizing hybrid sounds. You can hear additional audio examples on the companion website.

\section{Creative Applications}

\subsection{Interface}
For our model to be used in a studio production context, we have developed a user interface. This interface is a Max4Live patch which allows a direct integration into Ableton Live. In this section, we describe how it works and show some screen-shots.

To recall, we pass a (latent code, category) couple $(z,c)$ to the decoder of our CWAE to produce a spectrogram $\hat{x}$. Then the MCNN generates a .wav file from this spectrogram. However, the latent code $z$ is high dimensional (64 dimensions), so choosing a value for each parameter would be a long and complex process. To facilitate interactivity, we decided to use a Principal Components Analysis (PCA) which aim is to find the 3 most influential dimensions, thus reducing the complexity of the fine tuning process while ensuring a good diversity in sounds. From now on, we denote the PCA dimensions $P_1$, $P_2$ and $P_3$.

To generate sound through the interface, we provide controllers: First, we provide control over the values for $z$: an XY pad allows to control $P_1$ and $P_2$ and the 'Fine' knob provides control over $P_3$. Also, a selector allows the user to define the range of both the pad and the knob. Then, a menu allows the user to set a value for $c$ which comes down to selecting the type of sounds one wants to generate. Finally, the user can use the waveform visualizer to crop out remaining artifacts for example.

\subsection{Generation Process}
Every time a parameter value changes, a new sound is generated as follows. 
A python server is listening on a UDP port. This server contains the model and will be in charge of all the computation. When the user modifies the value of a dimension, the Max client sends a message via UDP. This message contains the values for $P_1$, $P_2$, $P_3$, and the category of the sound. When the server receives the message, it creates the associated latent code $z$ by computing the inverse PCA of $(P_1, P_2, P_3)$ and concatenate it with the conditioning vector. Then the server passes $(z,c)$ to the CWAE decoder which feeds a spectrogram to the MCNN. The obtained waveform is then exported to a WAV file, and its location is returned to the Max plugin. Finally, our plugin loads its buffer with the content of this file and displays it on the visualizer.

Our system can generate sounds with very low latency on CPU ($<$50ms delay between the change and the sound with a 2,6 GHz Intel Core i7). Once the sound is in the buffer, it can be played without any latency. A demonstration video is available on the companion website.

\subsection{Impact on creativity and music production}
We think that this system is a first approach towards a new way to design and compose drums. Indeed, it is a straightforward and efficient tool for everyone to organize and browse their sample library and design their drum sounds. Despite the parameters being autonomously learnt by the neural network, it is pretty intuitive to navigate in the latent space. 

Also, such a tool can be used to humanize programmed drums. It is often claimed that programmed electronic drums lack a human feeling. Indeed, when a real drummer plays, subtle variations give the rhythm a natural groove whereas programmed MIDI drum sequences can sound robotic and repetitive, leaving listeners bored.
There are common techniques to humanize MIDI drums such as varying velocities. By allowing the synthesis parameters to vary in a small given range, our system can be used to slightly modify the sound of a drum element throughout a loop. This could, for example, mimic a drummer who hits a snare at slightly different positions.

\section{Conclusion and Future Work}
We propose a first end-to-end system that allows intuitive drum sounds synthesis. The latent space learnt on the data provides intuitive controls over the sound. Our system is capable of real-time sound generation on CPU while ensuring a satisfying audio quality. Moreover, the interface we have developed is studio-ready and allows users to easily integrate it into one of the most used DAWs for electronic music. 
We identify two axes for improvement:
The first one is about the conditioning mechanism that should be more precise and powerful so that each category can clearly be distinguished from the others.
The other axis is about developing novel ways to interact with a large latent space to explore its full diversity.
Also, similarly to what is achieved on symbolic music \cite{latentconstraints2017,hadjeres2019variation}, we will investigate approaches that let the users specify the controls they want to shape the sounds. This would be an effortless way for novice sound designers to tune their drum sounds and create drum kits on purpose, rather than relying on existing ones. Also, to merge the computation server into the plugin is a required feature for the model to be even more accessible.

\bibliographystyle{iccc}
\bibliography{iccc}

\end{document}